# The Binary-Symmetric Parallel-Relay Network

Lawrence Ong, Sarah J. Johnson, and Christopher M. Kellett

**Abstract**

We present capacity results of the binary-symmetric parallel-relay network, where there is one source, one destination, and $K$ relays in parallel. We show that forwarding relays, where the relays merely transmit their received signals, achieve the capacity in two ways: with coded transmission at the source and a finite number of relays, or uncoded transmission at the source and a sufficiently large number of relays. On the other hand, decoding relays, where the relays decode the source message, re-encode, and forward it to the destination, achieve the capacity when the number of relays is small.

## 1 Introduction

In this paper, we present capacity results for the binary-symmetric parallel-relay (BSPR) network, where a source transmits data to a destination via relays (refer to Fig. 1).

We show that using a finite number of *forwarding relays*, which forward their received signals without any decoding and re-encoding, performs arbitrarily close to the capacity of the network as the number of relays increases. The asymptotic capacity results hold even when the source sends uncoded message bits. For networks with a small number[§] of relays, forwarding relays do not achieve the capacity, but *decoding relays*, where the relays decode the source message, re-encode, and forward it to the destination, do achieve the capacity.

The additive white Gaussian noise (AWGN) parallel-relay network was studied in [1–3] (and references therein). It has been shown that using coding at the source and no coding at the relays achieves the asymptotic capacity of the AWGN parallel-relay network [2] and the non-coherent multiple-input multiple-output (MIMO) relay network [4] as the number of relays increases to infinity. The source encodes the data and the relays merely scale their received signals and forward them. While these results hold only for large AWGN networks and have not been proven for the other networks, this paper considers the BSPR network and stronger results are obtained. To the best of our knowledge, the BSPR network has not previously been investigated. We will show that using coded transmission at the source and mere forwarding at the relays achieves

---

This work is supported by the Australian Research Council under grant DP0877258.

[§]The terms "small" and "medium" will be made more precise in the subsequent sections.

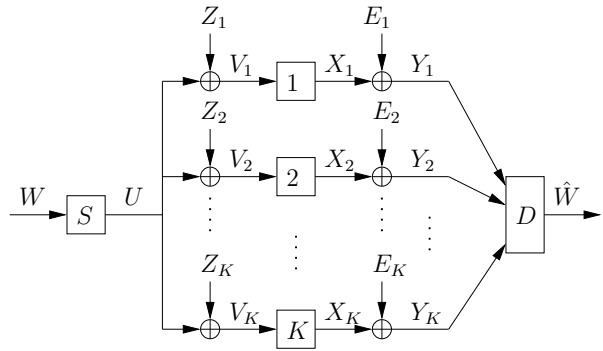

Figure 1: The BSPR network with $K$ relays

Table 1: Conditions under which various coding schemes achieve the capacity

| Coding schemes | Delay | Network size ($K+2$) |
|---|---|---|
| Uncoded transmission with forwarding relays | Two channel uses | Infinitely large |
| Coded transmission with forwarding relays | Infinitely long | Medium[§] to infinitely large |
| Coded transmission with decoding relays | Infinitely long | Small[§] |

arbitrarily close to the capacity of the BSPR network with a *finite* number of relays.

Uncoded transmission seldom achieves the capacities of noisy channels. However, it has been shown to be optimal (in terms of distortion measure) in the AWGN parallel-relay network with any number of relays, when the source messages are Gaussian random variables [5]. This suggests that when the source is *matched* to the channel, uncoded transmission could be optimal for other parallel-relay networks. For the BSPR network, we show that when the source (which outputs random bits) is matched to the binary channel, uncoded transmission achieves the capacity asymptotically when the number of relays, $K$, tends to infinity, but is suboptimal when $K$ is small.

The main observations of the paper are summarized in Table 1. It gives conditions under which different coding schemes achieve the capacity of the BSPR network. The delay is defined as the lag between the time a message bit is transmitted at the source and the time the message bit is decoded at the destination. The observations are deduced from the following theorems derived in this paper: Theorems 1 and 2 which give two upper bounds to the capacity of the BSPR network, Theorems 3 and 6 which give two achievable rate regions, Theorem 7 which gives the capacity for $K = 1$ and for $K > 1$ under certain conditions, and Theorems 4 and 5 which give asymptotic capacity results as $K$ increases. In some cases where none of the three coding schemes achieve the capacity, a hybrid scheme (Theorem 8) proves to be useful.

## 2 Channel Model and Notation

The BSPR network with $K$ relays is depicted in Fig. 1. The network has one source (node $S$), $K$ relays (nodes 1, 2, ..., $K$), and one destination (node $D$). A message $W$ is observed by the source and is to be communicated to the destination. $W$, $U$, $V_i$, $X_i$, $Y_i$, $E_i$, $Z_i$, $\forall i$, are binary variables.

At time $t$, the channel from the source to the $i$-th relay is a binary-symmetric channel (BSC) given by $V_i[t] = U[t] \oplus Z_i[t]$, where $U[t]$ is the signal transmitted by the source, $V_i[t]$ is the signal received by the $i$-th relay, $Z_i[t]$ is the random channel noise with $\Pr\{Z_i[t] = 1\} = p_{s,i}$, and $\oplus$ is modulo 2 addition. The probability $p_{s,i}$ is also known as the cross-over probability. Without loss of generality, we assume that $0 \leq p_{s,i} \leq \frac{1}{2}$. Similarly, at time $t$, the channel from relay $i$ to the destination is a BSC given by $Y_i[t] = X_i[t] \oplus E_i[t]$, where $X_i[t]$ is the signal transmitted by relay $i$, $Y_i[t]$ is the signal received at the destination, $E_i[t]$ is the random channel noise with $\Pr\{E_i[t] = 1\} = p_{i,d}$. Again, we assume that $0 \leq p_{i,d} \leq \frac{1}{2}$. We assume that all $Z_i[t]$ and $E_i[t]$ are independent.

Let $W$ be randomly and uniformly chosen from $\{0, 1, \ldots, 2^{nR}-1\}$. Consider $n$ channel uses from which the destination produces a source estimate $\hat{W}$. The capacity, $C$, is the supremum of all achievable rates $R$ at which $\hat{W}$ can be made to satisfy $P_e = \Pr\{\hat{W} \neq W\} < \epsilon$ for any $\epsilon > 0$.

## 3 Upper Bounds to Capacity

We first derive two upper bounds to the capacity.

**Theorem 1.** *Any achievable rate of the BSPR network must satisfy $R \leq 1$.*

Since the source can send at most 1 bit/channel use (through the binary $U$), the capacity is bounded by this rate. Though this capacity upper bound might seem loose at first sight, we will show that this bound is achievable (asymptotically) as $K$ increases. Next, we have a tighter upper bound:

**Theorem 2.** *Any achievable rate of the BSPR network must satisfy*

$$R \leq \min\left\{ I(U; V_1, V_2 \ldots, V_K), \sum_{i=1}^{K} I(X_i; Y_i) \right\}, \qquad (1)$$

*for some $p(u)p(x_1)\cdots p(x_K)$.*

Upper bounds of the type in Theorem 2 are often called cut-set upper bounds [6, Theorem 15.10.1]. The first term on the RHS of (1) is the maximum rate that information can transfer across the cut separating the source from the relays and the destination, and the second term, the cut separating the destination from the source and the relays.

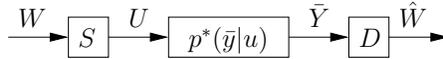

Figure 2: The equivalent point-to-point channel with forwarding relays, where $\bar{Y} = (Y_1, Y_2, \ldots, Y_K)$ and $p^*(\bar{y}|u)$ defined in (4)–(5)

## 4 Lower Bounds to Capacity

### 4.1 Coded Transmission with Forwarding Relays

We first consider relays which do not decode their received signals nor re-encode them.

**Definition 1.** *Relay $i$ is called a* forwarding relay *if it simply forwards its received signals: $X_i[t] = V_i[t-1]$, $\forall t$.*

Using only forwarding relays, the received signals at the destination can be re-written as:
$$Y_i = U \oplus Z_i \oplus E_i = U \oplus N_i, \quad i \in [1, K], \tag{2}$$

where $N_i = Z_i \oplus E_i$. We have dropped the time indices in the equation above as it is clear that the destination receives the noisy version of the source's transmission after two channel uses. We can easily show that

$$\Pr\{N_i = 1\} = p_{s,i}(1 - p_{i,d}) + (1 - p_{s,i})p_{i,d} \triangleq p_i, \tag{3}$$

which is the cross-over probability for the BSC from $U$ to $Y_i$.

We have thus turned the network in Fig. 1 with forwarding relays into a point-to-point channel from $U$ to $\bar{Y}$ as depicted in Fig. 2, where $\bar{Y} = (Y_1, Y_2, \ldots, Y_K) \in \{0,1\}^K$. The channel can be completely defined by the transition probability

$$p^*(\bar{y}|u) = \prod_{i=1}^{K} p^*(y_i|u), \tag{4}$$

where

$$p^*(y_i|u) = \begin{cases} 1 - p_i, & \text{if } y_i = u \\ p_i, & \text{otherwise} \end{cases}. \tag{5}$$

**Theorem 3.** *Consider the BSPR network depicted in Fig. 1. Rates up to the following are achievable:*

$$R_{\mathrm{F}} = I(U; Y_1, Y_2, \ldots, Y_K), \tag{6}$$

*for some distribution $p(u) \prod_{i=1}^{K} p^*(y_i|u)$, where $p^*(\cdot)$ is defined in (5) and (3).*

*Proof of Theorem 3.* Consider the point-to-point channel in Fig. 2. Theorem 3 follows from the channel coding theorem [6, Theorem 7.7.1]. □

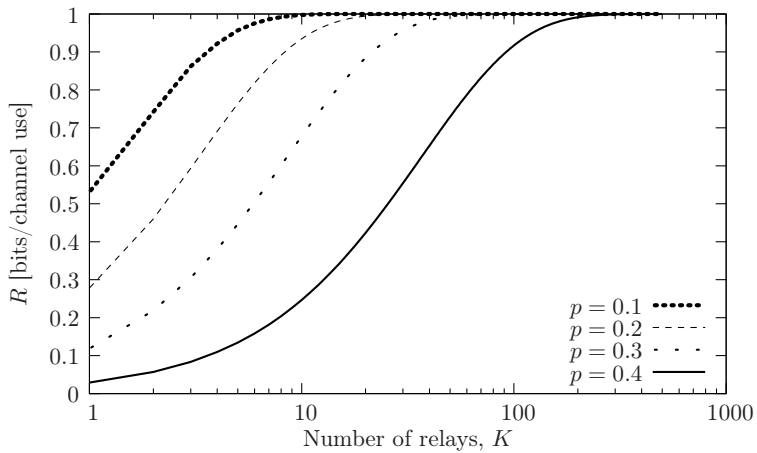

Figure 3: Achievable rates (with $P_e \to 0$) of coded transmission with forwarding relays vs. the number of relays ($p_i = p, \forall i$)

**Special case:** $p_i = p$, $\forall i \in [1, K]$

The computational complexity of the expression $I(U; Y_1, Y_2, \ldots, Y_K)$ in (6) is in the order of $\mathcal{O}(2^K)$. For the special case of $p_i = p, \forall i \in [1, K]$, the expression can be evaluated to the following:

**Lemma 1.** *Consider the BSPR network depicted in Fig. 1 and $p_i = p$, $\forall i \in [1, K]$. With forwarding relays, rates up to the following are achievable:*

$$R_\text{F} = 1 + Kp \log p + Kq \log q - \sum_{l=0}^{K-1} \binom{K-1}{l}(q^l p^{K-l} + q^{K-l} p^l) \log(q^l p^{K-l} + q^{K-l} p^l), \tag{7}$$

*where $q = 1 - p$.*

We compute achievable rates of coded transmission with forwarding relays for $p = 0.1, 0.2, 0.3,$ and $0.4$. Note that for $p = 0.5$, the channels are *randomized* and the capacity is zero. The results are shown in Fig. 3. At $p = 0.1$, we only need 16 relays or more to achieve 0.9999 bits/channel use. At $p = 0.4$, we need 387 relays or more to achieve 0.9999 bits/channel use.

Note that $I(U; Y_1, Y_2, \ldots, Y_{K+1}) > I(U; Y_1, Y_2, \ldots, Y_K)$ as the noise in the links are independent. So, having more relays in the network increases the achievable rate. From numerical computations, we can show that for any $p < \frac{1}{2}$, there is a finite $K(p) > 0$ for which the capacity upper bound of 1 bit/channel use (rounded to some significant figures) is achievable with $K(p)$ or more forwarding relays.

## 4.2 Uncoded Transmission with Forwarding Relays

Now, we derive achievable rates for the case where the source sends uncoded bits ($U = W$) and the relays do simple forwarding ($X_i = V_i$). This means the source sends one bit of raw information per channel use. i.e., at the rate $R = 1$.

**Remark 1.** *In the previous section, we derived different achievable rates for networks with different numbers of relays. In this section, we fix the transmission rate at 1 bit/channel use (which is a capacity upper bound), and analyze under what condition this rate is achievable.*

As forwarding relays are used, we have the equivalent channel in Fig. 2 with $U = W$. At the destination, the received signals are:

$$Y_i = W \oplus N_i, \quad i \in [1, K], \tag{8}$$

where $N_i = Z_i \oplus E_i$ and $\Pr\{N_i = 1\} = p_i$ as defined in (3).

Let $\bar{y} = (y_1, y_2, \ldots, y_K)$ be the received signals at the destination. The optimal decision decoding rule, which minimizes the error probability, is:

$$\hat{w} = \begin{cases} 0, \text{if } \Pr\{\bar{Y} = \bar{y}|W = 0\} \geq \Pr\{\bar{Y} = \bar{y}|W = 1\} \\ 1, \text{otherwise} \end{cases}. \tag{9}$$

**Special case:** $p_i = p,\ \forall i \in [1, K]$

First, consider the case $p_i = p, \forall i \in [1, K]$. The decision rule in (9) becomes

$$(1-p)^{\mathbf{0}(\bar{y})} p^{K-\mathbf{0}(\bar{y})} \underset{0}{\overset{1}{\gtreqless}} (1-p)^{K-\mathbf{0}(\bar{y})} p^{\mathbf{0}(\bar{y})}, \tag{10}$$

where $\mathbf{0}(\bar{y})$ is the number of 0's in $\bar{y}$. As we can show that $0 \leq p \leq \frac{1}{2}$, the optimal decoding function is:

$$\hat{W} = g(\bar{Y}) = \begin{cases} 0, & \text{if } \mathbf{0}(\bar{Y}) \geq \frac{K}{2} \\ 1, & \text{otherwise} \end{cases}. \tag{11}$$

Having defined the decoding function, we derive the error probability.

$$\begin{aligned} P_e &= \Pr\{\hat{W} \neq W\} \\ &= \frac{1}{2} \Pr\left\{\mathbf{0}(\bar{Y}) < \frac{K}{2} \Big| W = 0\right\} + \frac{1}{2} \Pr\left\{\mathbf{0}(\bar{Y}) \geq \frac{K}{2} \Big| W = 1\right\} \\ &= \Pr\left\{\mathbf{0}(\bar{Y}) \geq \frac{K}{2} \Big| W = 1\right\} \\ &= \Pr\left\{\sum_{i=1}^{K} N_i \geq \frac{K}{2}\right\} \\ &= \Pr\left\{\frac{1}{K} \sum_{i=1}^{K} N_i - p \geq \frac{1}{2} - p\right\} \\ &\leq \exp\left[-2K\left(\frac{1}{2} - p\right)^2\right], \end{aligned} \tag{12}$$

where the inequality in (12) is due to Hoeffding [7, Theorem 2] if $\frac{1}{2} - p > 0$.

This gives us the following theorem.

**Theorem 4.** *Consider the BSPR network in Fig. 1 and $p_i = p < \frac{1}{2}$, $\forall i \in [1, K]$. Using uncoded transmission at the source and forwarding relays, the asymptotic capacity of 1 bit/channel is achievable with a sufficiently large number of relays.*

*Proof of Theorem 4.* By sending uncoded bits at the rate 1 bits/channel use, we know from (12) that the error probability can be bounded by $P_e \leq \exp(-K\delta)$, where $\delta = 2\left(\frac{1}{2} - p\right)^2 > 0$. So, for any $0 \leq p < \frac{1}{2}$ and $\epsilon > 0$, we can select $K > \frac{1}{\delta} \ln \frac{1}{\epsilon}$ such that $P_e < \epsilon$. Since 1 bit/channel use is an upper bound to the capacity, we have Theorem 4. □

**General case: possibly different $p_i$, $1 \leq i \leq K$**

For the general case where not all $p_i$ are equal, the decoding rule in (11) is not optimal. However, we can modify the received signals at the destination and still use the decoding rule in (11) to obtain asymptotic capacity results.

**Theorem 5.** *Consider the BSPR network depicted in Fig. 1. We use uncoded transmission at the source and forwarding relays. Let a subset of $M$ relays be $\mathcal{M} = \{m_1, m_2, \ldots, m_M\} \subseteq [1, K]$ and let $p_{\mathrm{MAX}(\mathcal{M})} = \max_{m_i \in \mathcal{M}} p_{m_i}$. If there exists an $\mathcal{M}$ for each $K$ such that*

$$|\mathcal{M}| \left(\frac{1}{2} - p_{\mathrm{MAX}(\mathcal{M})}\right)^2 \to \infty \quad \text{as} \quad K \to \infty, \tag{13}$$

*then the asymptotic capacity of 1 bit/channel use is achievable with a sufficiently large number of relays.*

The proof of the above theorem follows from Theorem 4 and is omitted because of space limitations.

**Remark 2.** *The condition in (13) is not unreasonable as long as when the number of relays increases, the number of "bad" channels (with cross-over probabilities $p_i$ close to $\frac{1}{2}$) can be kept at or below a certain fraction of $K$. A trivial example of a network satisfying this condition is one in which as $K$ increases, the maximum cross-over probability is unchanged at some $p_{\mathrm{MAX}([1,K])} = p_{\mathrm{const}} < \frac{1}{2}$.*

### 4.3 Coded Transmission with Decoding Relays

In this section, we investigate if better results can be obtained if we let the relays decode the source messages, re-encode, and forward them to the destination. We now state the rate achievable with decoding at some relays.

**Theorem 6.** *Consider the BSPR network depicted in Fig. 1. Rates up to the following are achievable:*

$$R_{\mathrm{D}} = \min\left\{\left\{I(U; V_{m_i}) : \forall m_i \in \mathcal{M}\right\}, \sum_{i=1}^{M} I(X_{m_i}; Y_{m_i})\right\} \tag{14}$$

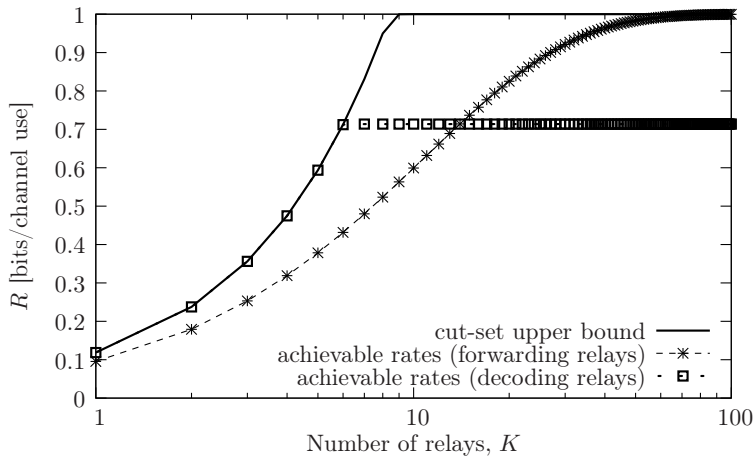

Figure 4: Cut-set upper bound and achievable rates with coding at the source, $p_s = 0.05$, $p_d = 0.3$

for some $\mathcal{M} = \{m_1, \ldots, m_M\} \subseteq [1, K]$, and some distribution $p(u) \prod_{i=1}^{M} p(v_{m_i}|u)$ $p(x_{m_1})p(x_{m_2}) \cdots p(x_{m_M}) \prod_{j=1}^{M} p(y_{m_j}|x_{m_j})$.

*Sketch of Proof for Theorem 6:* The first term on the RHS of (14) is the rate at which all relays in $\mathcal{M}$ can decode the source message, and the second term, the rate at which the destination can decode the messages from the relays $\mathcal{M}$. The complete proof is omitted because of space limitations.

**Special case:** $p_{s,i} = p_s$ **and** $p_{i,d} = p_d$**,** $\forall i \in [1, K]$

Since the uniform input distribution is optimal for the BSC, setting $U$ and $X_{m_i}, \forall m_i \in \mathcal{M}$, to be independently and uniformly distributed simultaneously maximizes all the mutual information terms in (14). This means the mutual information terms in (14) are evaluated as follows:

$$I(U, V_{m_i}) = H(V_{m_i}) - H(V_{m_i}|U) = 1 - H(p_{s,m_i}), \tag{15}$$

$$\sum_{i=1}^{M} I(X_{m_i}; Y_{m_i}) = \sum_{i=1}^{M} (1 - H(p_{m_i,d})) = M - \sum_{i=1}^{M} H(p_{m_i,d}). \tag{16}$$

When $p_{s,i} = p_s$, $\forall i$, we have $1 - H(p_{s,i}) = 1 - H(p_{s,j})$, $\forall i, j$. So, selecting any $\mathcal{M}$ will not affect $\min_{m_i \in \mathcal{M}} \{1 - H(p_{s,m_i})\}$ in (14). In addition, adding more relays into the set $\mathcal{M}$ can never decrease (16) as $H(p_{m_i,d}) \leq 1$. Hence, it is always optimal to set $\mathcal{M} = [1, K]$. Doing this, the rate in Theorem 6 for the case where $p_{s,i} = p_s$ and $p_{i,d} = p_d$, $\forall i \in [1, K]$ becomes

$$R_D = \min\{1 - H(p_s), K(1 - H(p_d))\}. \tag{17}$$

Now, we compare the achievable rates of coded transmission with forwarding relays and with decoding relays to the cut-set upper bound. Fig. 4 shows achievable rates with the following parameter values: $p_s = 0.05$, $p_d = 0.3$. Decoding relays achieve a higher transmission rate compared to forwarding relays when the number of relays is small, but lower when the number of relays increases. From (17), we see that when $K \geq \frac{1-H(p_s)}{1-H(p_d)}$, the achievable rate region of decoding relays is fixed at $R \leq 1 - H(p_s) < 1$, which is sub-optimal when $K$ grows.

**Decoding relays achieve the capacity under certain conditions**

From Fig. 4, we see that decoding relays achieve the capacity when $1 \leq K \leq 6$. Here we characterize the exact conditions for which decoding relays achieve the capacity.

**Theorem 7.** *Consider the BSPR network depicted in Fig. 1. The capacity is known for the following conditions:*

1. *If K=1, then the capacity is*

$$C = \max_{p(u)p(x_1)} \min \{I(U;V_1), I(X_1;Y_1)\}. \tag{18}$$

2. *If K>1, and if*

$$\sum_{j=1}^{K} \max_{p(x_j)} I(X_j;Y_j) \leq \max_{p(u)} I(U;V_i), \ \forall i \in [1,K], \tag{19}$$

*then the capacity is*

$$C = \sum_{i=1}^{K} \max_{p(x_i)} I(X_i;Y_i). \tag{20}$$

*Proof.* From Theorem 2, the cut-set upper bound is

$$R \leq \max_{p(u)p(x_1)\cdots p(x_K)} \min \left\{ I(U;V_1,\ldots,V_K), \sum_{i=1}^{K} I(X_i;Y_i) \right\}.$$

From Theorem 6, taking $\mathcal{M} = [1,K]$, the achievable rate region of decoding relays is

$$R \leq \max_{p(u)p(x_1)\cdots p(x_K)} \min \left\{ \{I(U;V_i) : \forall i\}, \sum_{i=1}^{K} I(X_i;Y_i) \right\}.$$

Obviously, when $K = 1$, the two regions match.

For $K > 1$, if (19) is satisfied, the achievable rate region of decoding relays reduces to (20). Since $I(U;V_1,\ldots,V_K) \geq I(U;V_i), \forall i$, we have

$$\max_{p(u)} I(U;V_1,V_2\ldots,V_K) \geq \sum_{i=1}^{K} \max_{p(x_1)} I(X_i;Y_i), \tag{21}$$

and the upper bound also reduces to (20). □

**Remark 3.** *Note that condition (19) means that the sum of capacities of all $K$ channels from the relays to the destination is smaller than the capacity of the channel from the source to each relay. This condition is likely to hold for small $K$ and when the channels from the relays to the destination are noisy. As the LHS is the sum of mutual information terms for $K$ channels, the condition is unlikely to hold for large $K$.*

### 4.4 A Hybrid Coding Scheme

Using coded transmission and having only the first $M$, $M \leq K$, of the relays forwarding, we can achieve rates up to $R_\text{F} = I(U; Y_1, Y_2, \ldots, Y_M)$. Note that using more relays always increases $R_\text{F}$. However, using forwarding relays, the noise from the source-to-relays links propagates to the relays-to-destination links. This can be rectified by having these $M$ relays decoding and re-encoding the source messages. This increases the rate at which the destination can decode to $\sum_{i=1}^{M} I(X_i, Y_i) \geq I(U; Y_1, Y_2, \ldots, Y_M) = R_\text{F}$. However, forcing these relays to decode adds an additional constraint of $1 - H\left(\max_{1 \leq i \leq M} p_{s,i}\right)$ on the rate.

We propose a hybrid coding scheme in which some relays decode and re-encode the messages, and the rest of the relays forward their received signals. Denote the set of decoding relays by $\mathcal{D}$ and the set of forwarding relays by $\mathcal{F} = \{1, 2, \ldots, K\} \setminus \mathcal{D}$. Since all relays in the set $\mathcal{D}$ need to decode the messages, the rate is constrained by $1 - H\left(\max_{i \in \mathcal{D}} p_{s,i}\right)$. As all relays in $\mathcal{D}$ fully decode the source message and re-encode, we can think of the BSPR network as a point-to-point MIMO channel from $(U, X_\mathcal{D})$ to $(Y_1, Y_2, \ldots, Y_K)$, where $U$ and $X_i$, for all $i \in \mathcal{D}$, are statistically independent. The destination node can decode the messages at the rate $I(U, X_\mathcal{D}; Y_1, Y_2, \ldots, Y_K) = I(U; Y_\mathcal{F}) + I(X_\mathcal{D}; Y_\mathcal{D}) = I(U; Y_\mathcal{F}) + \sum_{i \in \mathcal{D}} I(X_i, Y_i)$. So, we have the following theorem:

**Theorem 8.** *Consider the BSPR network depicted in Fig. 1. Rates up to the following are achievable:*

$$R_\text{H} = \left\{ 1 - H\left(\max_{i \in \mathcal{D}} p_{s,i}\right), I(U; Y_\mathcal{F}) + \sum_{i \in \mathcal{D}} \left(1 - H(p_{i,d})\right) \right\},$$

*for any $\mathcal{D} \subseteq \{1, 2, \ldots, K\}$ and $\mathcal{F} = \{1, 2, \ldots, K\} \setminus \mathcal{D}$, where $H\left(\max_{i \in \mathcal{D}} p_{s,i}\right) = 0$ if $|\mathcal{D}| = 0$, and $p(u)$ is uniform.*

Determining the optimal set of $\mathcal{D}$ is a hard combinatorial problem. A rule of thumb is to include relays with low $p_{s,i}$ in $\mathcal{D}$, and relays with high $p_{s,i}$ in $\mathcal{F}$.

Fig. 5 shows rates achievable using the hybrid scheme, all forwarding relays, and only decoding relays for a BSPR network with $K = 8$ relays, $p_{s,i} = \frac{0.1i}{K}$, and $p_{i,d} = 0.3$ for all $i \in [1, K]$. We vary the number of relays to be included in the set of decoding relays $\mathcal{D}$. Setting all relays to perform forwarding, we achieve $R_F = 0.52$ (i.e., the hybrid scheme when $|\mathcal{D}| = 0$). Using only decoding relays, we achieve $R_D = 0.62$ when six of the eight relays perform decoding and re-encoding (and the rest of the relays are unused). Using a combination of forwarding and decoding relays, we can achieve a significantly higher rate of $R_H = 0.71$ using four decoding relays and four forwarding relays. For each $|\mathcal{D}|$, the optimal relay selection is done by the brute force search.

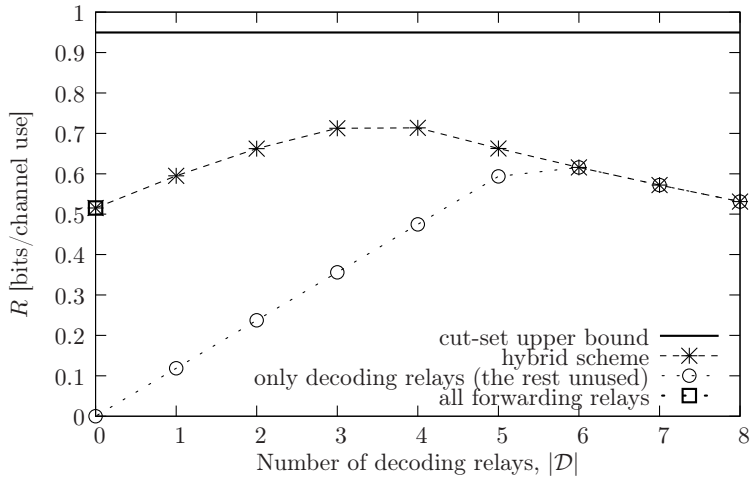

Figure 5: Comparing the hybrid coding scheme to the other coding schemes and the cut-set upper bound, $K = 8$, $p_{s,i} = \frac{0.1i}{K}$, $p_{i,d} = 0.3$, $\forall i \in [1, K]$

## 5 Reflection

The first three schemes discussed in this paper are capable of achieving the capacity of the BSPR network under different conditions, as summarized in Table 1.

With coded transmission, decoding relays achieve the capacities of networks with one relay, and networks with more relays if the sum of capacities of all the channels from the relays to the destination is smaller than the capacity of the channel from the source to each relay (this condition is likely to hold only when the number of relays is small).

With coded transmission, forwarding relays achieve 1 bit/channel use (rounded to some significant figures) for networks with more than a certain number of relays. We showed this by numerical calculations in Fig. 3.

Using coded transmission, a sufficiently long code length is required to drive the error probability to zero. This necessarily incurs a large delay. With uncoded transmission, on the other hand, decoding is almost "instantaneous" – the destination can decode the source bit after two channel uses. With uncoded transmission, message bits are sent at 1 bits/channel use (capacity upper bound). To drive the error probability to zero, however, a sufficiently large number of relays is required.

When neither decoding relays nor forwarding relays achieve the capacity upper bound, we proposed a hybrid scheme that can achieve rates significantly higher than those achievable using only forwarding or only decoding relays.

# References

[1] B. Schein and R. Gallager, "The Gaussian parallel relay network," in *Proc. IEEE Int. Symposium on Inf. Theory (ISIT)*, Sorrento, Italy, Jun. 25-30 2000, p. 22.

[2] M. Gastpar and M. Vetterli, "On the capacity of large Gaussian relay networks," *IEEE Trans. Inf. Theory*, vol. 51, no. 3, pp. 765–779, Mar. 2005.

[3] S. S. C. Rezaei, S. O. Gharan, and A. K. Khadani, "A new achievable rate for the Gaussian parallel relay channel," in *Proc. IEEE Int. Symposium on Inf. Theory (ISIT)*, Seoul, Korea, Jun. 28-Jul. 3 2009, pp. 194–198.

[4] H. Bölcskei, R. U. Nabar, O. Oyman, and A. J. Paulraj, "Capacity scaling laws in MIMO relay networks," *IEEE Trans. Wireless Commun.*, vol. 5, no. 6, pp. 1433–1444, Jun. 2006.

[5] M. Gastpar, "Uncoded transmission is exactly optimal for a simple Gaussian "sensor" network," *IEEE Trans. Inf. Theory*, vol. 54, no. 11, pp. 5247–5251, Nov. 2008.

[6] T. M. Cover and J. A. Thomas, *Elements of Information Theory*, 2nd ed. Wiley-Interscience, 2006.

[7] W. Hoeffding, "Probability inequalities for sums of bounded random variables," *Journal of the American Statistical Association*, vol. 58, no. 301, pp. 13–30, Mar. 1963.